\documentclass[cmp]{svjour}

\usepackage{amsmath}
\usepackage{amssymb}
\usepackage{bigints}
\usepackage{bm}
\usepackage{colortbl}
\usepackage{graphicx}
\usepackage{hyperref}
\usepackage{microtype}
\usepackage{multirow}
\usepackage{physics}
\usepackage{relsize}
\usepackage{times}
\usepackage{xcolor}

\newcommand*\dif{\mathop{}\!\mathrm{d}}

\newcommand{\aalpha}{\bm \alpha}
\newcommand{\kk}{{\mathbf k}}
\newcommand{\vv}{{\mathbf v}}

\newcommand{\zz}{{\mathbf 0}}

\journalname{Communications in Mathematical Physics}

\begin{document}

\title{Exceptional Lattice Green's Functions}
\titlerunning{Exceptional Lattice Green's Functions}

\author{Samuel Savitz \and Marcus Bintz}
\institute{Institute of Quantum Information and Matter, Department of Physics, California Institute of Technology, Pasadena,~California~91125,~USA\\
  \email{Sam@Savitz.org}}
\authorrunning{S.~Savitz and M.~Bintz}

\date{Uploaded: October 25\textsuperscript {th}, 2017}

\maketitle
\begin{abstract}
  The three exceptional lattices, $E_6$, $E_7$, and $E_8$, have attracted much attention due to their anomalously dense and symmetric structures which are of critical importance in modern theoretical physics.  Here, we study the electronic band structure of a single spinless quantum particle hopping between their nearest-neighbor lattice points in the tight-binding limit.  Using Markov chain Monte Carlo methods, we numerically sample their lattice Green's functions, densities of states, and random walk return probabilities.  We find and tabulate a plethora of Van Hove singularities in the densities of states, including degenerate ones in $E_6$ and $E_7$.  Finally, we use brute force enumeration to count the number of distinct closed walks of length up to eight, which gives the first eight moments of the densities of states.
\end{abstract}

\section{Introduction}

Lattices are among the most common spatial structures arising in physics.  In particular, the energy minimizing process of crystallization frequently leads to the emergent generation of lattice structures.  Through Bloch's theorem~\cite{Bloch29}, the electronic structure, thermodynamics, and material properties of metals are known to be heavily influenced by their lattice.  Here, we study one of the simplest models exhibiting this effect, a tight-binding model with nearest-neighbor hopping.

The mathematical theory of lattices is very mature, at least when the dimensionality is not too large.  The irreducible simply laced lattices (see below) are particularly elegant, physically relevant, and well understood.  They exhibit one of the most important instances of the \emph {ADE classification}, which has roots in Plato's enumeration of the five convex regular polyhedra~\cite {McKay80}.  The ADE classification is best known in association with the simply laced Lie groups, as established by Wilhelm Killing in 1888~\cite {Killing1888}, and has been found to recur surprisingly often throughout mathematics, including rational singularities~\cite{Arnold85}, quivers~\cite{Gabriel72}, and conformal field theories~\cite{Cappelli1987}.  In each instance, one finds two infinite $A_d$ and $D_d$ families and three exceptional structures, $E_6$, $E_7$, and $E_8$.  The study of the ubiquitous appearance of this mathematical motif has been referred to as ``ADE-ology''~\cite{Cappelli:2010}.

The exceptional lattices are particularly familiar to physicists owing to their key applications in string theory~\cite{Gross85}. From a lower-energy perspective, these lattices possess remarkable density and symmetry, and so it is reasonable to expect that some hypothetical \mbox {six-,} \mbox {seven-,} and eight-dimensional metals would crystallize according to their structure~\cite {Savitz17}.  We therefore use these three exceptional lattices as the setting for our tight-binding model.

Using Monte Carlo techniques, we numerically evaluate the lattice Green's functions and densities of states for a spinless electron hopping on the exceptional lattices. Notably, these densities are highly skewed, with a particularly long tail as one approaches the ground state.  We give asymptotically exact expressions for their behavior at both extrema of their energy bands.  Additionally, the densities exhibit a number of non-smooth points known as Van Hove singularities~\cite{VanHove} which we also tabulate. 

There are close connections between these electronic properties and certain combinatorics problems from the field of random walks.  In particular, the moments of the densities of states are given by the numbers of distinct closed walks on the corresponding lattice.  We therefore present a computational enumeration of these walks of length up to eight, which we hope will be extended.  Finally, the lattice Green's functions can be used to calculate the \emph {return probabilities} that a random walker will return to the origin at any time during its infinite walk, which we numerically evaluate to at least three significant figures.

\section{Background}
\subsection{The ADE Lattices} \label {ADE}

A \emph {lattice} $\Lambda$ is a non-dense set of points in $\mathbb R^N$ which is closed under vector addition.  All lattices are isomorphic to $\mathbb Z^d$ for some $d \le N$, which is called the \emph {dimensionality} of the lattice, and contain the origin: $\zz \in \Lambda$.  One frequently imagines placing open $N$-balls of uniform radius on the lattice points.  The largest radius which leads to no point in $\mathbb R^N$ being inside of more than one ball is called the \emph {packing radius}, which is half the length of the shortest non-zero vector in $\Lambda$.  For the sake of convenience, we will scale all of our lattices such that the packing radius is $\sqrt 2$, leading to \emph {minimal vectors} of length $\sqrt 8$.  The number of minimal vectors gives the count of distinct hyperspheres in contact with any given central hypersphere, and is therefore called the \emph {kissing number} of the lattice, denoted by $\tau$.

The \emph {simply laced} lattices with no vectors of length $\sqrt {8} < |\vv| < 4$ are particularly important in math and physics due to their high density and symmetry.  Their minimal vectors $\aalpha_1, \ldots, \aalpha_\tau$ are called \emph {roots}.  One can show that the angle between any two roots as measured from the origin must be $0$, $\pi/3$, $\pi/2$, $2\pi/3$, or $\pi$.  If a lattice can not be written as the direct sum of two orthogonal sublattices, it is called \emph {irreducible}.  The irreducible simply laced lattices fall under an ADE classification, consisting of two infinite families and three \emph {exceptional lattices}.

For all positive integers $d$, the irreducible simply laced $A_d$ lattice has $\tau = d^2 + d$ minimal vectors most conveniently embedded in $\mathbb R^{d + 1}$ as all permutations of $(-2, 2, 0, \ldots, 0)$.  $A_1$ is the unique linear lattice, $A_2$ is hexagonal, and $A_3$ is the face-centered cubic lattice.

For $d \ge 4$ the face-centered hypercubic lattice $D_d$ contains $\tau = 2d^2 - 2d$ minimal vectors embedded in $\mathbb R^d$ as all permutations of $(\pm 2, \pm 2, 0, \ldots, 0)$.  $D_d$ contains $A_{d - 1}$, and $D_3$ is isomorphic to $A_3$.

Besides these two infinite families, there are also three exceptional irreducible simply laced lattices, $E_6$, $E_7$, and $E_8$, on which we will focus our attention.  $E_8$ has $\tau = 240$ minimal vectors given by all permutations of $(\pm 2, \pm 2, 0, 0, 0, 0, 0, 0)$ and those of $(\pm 1, \pm 1, \pm 1, \pm 1, \pm 1, \pm 1, \pm 1, \pm 1)$ with an even number of negative signs.  $E_8$ can also be constructed by superimposing two copies of $D_8$ according to $E_8 = D_8 \cup \left( D_8 + (1, 1, 1, 1, 1, 1, 1, 1) \right)$.

$E_6$ can be conveniently constructed as the sublattice of $E_8$ in the hyperplane defined by having the last three coordinates all equal.  Likewise, $E_7$ is the sublattice in the hyperplane with coordinates summing to zero.  Their kissing numbers are 72 and 126, respectively.  $E_5$ can be considered to be isomorphic to $D_5$.

The densest known sphere packings in dimensions one through eight are the lattices $A_1$, $A_2$, $A_3$, $D_4$, $D_5$, $E_6$, $E_7$, and $E_8$.  These are precisely the \emph {laminated lattices} of their respective dimensionality, which informally means that they can be constructed by ``stacking'' copied layers of the previous lattice in a maximally dense way.  This requires placing the second layer above the \emph {deepest holes} of the base lattice, \emph {i.e.}~those points in the span of the lattice which are maximally distant from all lattice points.  The $E_8$ case is degenerate in the sense that $D_8$ has holes large enough to accommodate an entire additional lattice point without requiring any lateral shift.  $E_8$ was recently proven to have both optimal packing density and kissing number~\cite {Viazovska17}.  The interested reader is referred to~\cite {ConwaySloane93} for a much more complete exposition regarding lattices.

In physics, the exceptional Lie algebras usually appear in their continuous Lie group form, but the exceptional lattices are sometimes also important. The $E_8$ lattice in particular plays a key role in the bosonic construction of the heterotic string through the toroidal compactifaction  onto $E_8 \times E_8$. All three exceptional lattices can similarly be used to formulate various supergravity theories in lower dimensions~\cite{Lerchie89}.  In a separate context, the ADE lattices appear in the integrable field theories that describe the deformation of certain conformal field theories away from criticality. When these deformations are affine Toda theories, the mass spectrum is proportional to the radii of the Gosset circles formed by projecting the roots onto a Coxeter plane~\cite{Mussardo2010,Kostant2010,Borthwick11}. In particular, the one-dimensional quantum Ising chain near criticality has been experimentally demonstrated to possess at least the first two quasiparticle excitations predicted by the $E_8$ Toda theory~\cite{Coldea2010}. 

\subsection{Tight-Binding Models with Nearest-Neighbor Hopping}

Here we study a single-particle tight-binding model with nearest-neighbor hopping, wherein a quantum wavefunction $\ket \psi$ assigns a complex amplitude $\braket{\vv}{\psi}$ to each exceptional lattice point $\vv$, defined by the discrete Laplacian Hamiltonian
\begin {equation}
  H = -\sum_{\vv \in \Lambda} \sum_{j = 1}^\tau \ket{\vv + \aalpha_j} \bra \vv.
\end {equation}
The unnormalized eigenstates of this Hamiltonian are plane waves labeled by their crystallographic momentum $\kk$: $\braket{\vv}{\kk} = e^{i \vv \cdot \kk}$, and $H \ket \kk = \epsilon(\kk) \ket \kk$.  The \emph {electronic band structure} or \emph {dispersion relation} $\epsilon(\kk)$ can be thought of as the Fourier transform of negative Dirac deltas arranged on the lattice's minimal vectors/roots:
\begin {equation}
  \epsilon(\kk) = -\sum_{j = 1}^\tau e^{i \aalpha_j \cdot \kk}.
\end {equation}

\subsection {Lattice Green's Functions}

Given a model of this form, one can define the \emph {on-site lattice Green's function} as
\begin {equation}
  G(\epsilon) = \frac 1 V \int \frac {\dif^d \kk} {\epsilon - \epsilon(\kk) - i0^+},
\end {equation}
where $\kk$ varies over a periodic unit cell or \emph {Brillouin zone} of $\epsilon(\kk)$ with volume $V$.  The mathematics necessary to solve for the $A_2$ hexagonal lattice Green's function was developed by Ramanujan in one of his theories of elliptic functions to alternative bases~\cite{Ramanujan} and introduced to physics by~\cite{Horiguchi72}.  The real and imaginary parts of the lattice Green's function determine each other through the Kramers--Kronig relation~\cite{King09}:
\begin {subequations} \label {KramersKronig}
\begin {equation}
  \Re{G(\epsilon)} = \frac 1 \pi \bigintssss_{\epsilon_\mathrm {min}}^{\epsilon_\mathrm {max}} \frac {\Im{G(\epsilon')}} {\epsilon - \epsilon'} \dif \epsilon',
\end {equation}
and
\begin {equation}
  \Im{G(\epsilon)} = \frac 1 \pi \bigintssss_{-\infty}^{\infty} \frac {\Re{G(\epsilon')}} {\epsilon' - \epsilon} \dif \epsilon'.
\end {equation}
\end {subequations}
Their signs and normalization, as well as the those of the energy $\epsilon$, are subject to variable arbitrary conventions.

\subsection {Densities of States}

The \emph {density of states} is of central importance to the determination of electronic properties and is defined by
\begin {equation}
	\rho(\epsilon) = \frac 1 V \int \delta(\epsilon(\kk) - \epsilon) \dif^d \kk.
\end {equation}
Lattice densities of states are compactly supported on $\left( \epsilon_\mathrm {min}, \epsilon_\mathrm {max} \right) $, are normalized such that
\begin {equation}
  \int_{\epsilon_\mathrm {min}}^{\epsilon_\mathrm {max}} \rho(\epsilon) \dif \epsilon = 1,
\end {equation}
and can be expressed in terms of the imaginary parts of lattice Green's functions: $\rho(\epsilon) = \Im {G(\epsilon)}/\pi$.  The ground state corresponds to all of the phases being fully aligned and has energy $\epsilon_\mathrm {min} = -\tau$.  In one dimension, the maximum energy state clearly alternates signs on adjacent lattice sites, giving $\epsilon_\mathrm {max} = -\epsilon_\mathrm {min} = \tau$, but this is no longer necessarily true in higher dimensions (which can support lattices other than the simple cubic $\mathbb Z^d$), and finding the momentum which maximizes $\epsilon(\kk)$ is a non-trivial task.

Many mathematical approaches have been developed for evaluating lattice Green's functions, including contour integrals~\cite{Ray14}, hypergeometric functions and Calabi--Yau differential equations~\cite{Guttmann10}, holonomic functions~\cite{Koutschan13,Zenine15,HassaniEtAl16}, and Chebyschev polynomials~\cite{Loh17}.  Here we use the Metropolis--Hastings algorithm~\cite{Hastings70} to numerically evaluate the densities of states with small relative errors even near the energy extrema.

\subsection{Van Hove Singularities}

As noted by Van Hove, Morse theory says that the periodicity of $\epsilon(\kk)$ implies the existence of many \emph {stationary} or \emph {critical points} of $\epsilon(\kk)$, satisfying $\grad_\kk \epsilon(\kk) = \zz$~\cite {VanHove}.  These lead to non-smooth \emph {Van Hove singularities} in the densities of states.  At each Van Hove singularity $\kk_0$, the \emph {Hessian}
\begin {equation}
  \left. \frac {\partial^2 \epsilon(\kk)} {\partial \kk_a \partial \kk_b} \right|_{\kk_0}, \label {Hessian}
\end {equation}
has $n_\downarrow$ negative and $n_\uparrow$ positive eigenvalues.  $n_\downarrow$ is called the \emph {index} of the singularity.  We present what we conjecture to be a complete enumeration of our exceptional lattice hopping model's Van Hove singularities.

\emph {Quadratic critical points}, where $n_\downarrow + n_\uparrow = d$, give singularities of order $d/2 - 1$.  On the other hand, $E_6$ and $E_7$ also exhibit \emph {degenerate critical points}, for which $n_\downarrow + n_\uparrow < d$.  These lead to stronger singularities in the densities of states and require further investigation.

The Van Hove singularities corresponding to the tails of the densities of states $\epsilon_\mathrm {min}$ and $\epsilon_\mathrm {max}$, and particularly the ratio of their energies, $\gamma = -\epsilon_\mathrm {min}/\epsilon_\mathrm {max}$, called the \emph {$\gamma$ skewness} or \emph {extrema imbalance}, are important for the energetic stabilization of crystalline structures in the Barkan--Diamant--Lifshitz model~\cite{BDL,Savitz17}.  We hope to elaborate on this further in future work.  Additionally, we derive exact expressions for the asymptotic behavior of the densities of states as one approaches these extrema.

\subsection{Closed Walk Counting}

The calculation of the moments of the densities of states distributions turns out to be equivalent to the combinatorial enumeration of closed walks of length $n$ on the lattice, $W_n$~\cite{MontrollWeiss65}:
\begin {equation}
	\frac 1 V \int \epsilon(\kk)^n \dif^d \kk = \int \epsilon^n \rho(\epsilon) \dif \epsilon = (-1)^n W_n. \label{GeneratingFunction}
\end {equation}
One can express the walk counts using multinomial coefficients:
\begin {equation}
	W_n = \mathlarger{\mathlarger{\mathlarger{\sum}}}_{c_1, \ldots, c_\tau \in \mathbb N^0} \delta \left( n - \sum_{j = 1}^\tau c_j \right) \delta^d \left( \sum_{j = 1}^\tau c_j \aalpha_j \right) \frac {n!} {\prod_{j = 1}^\tau c_j!},
\end {equation}
although this is not an efficient way to evaluate them.

We computationally enumerate $W_n$ for $n = 0 \ldots 8$ exactly.  Given sufficient terms, one could ``experimentally'' find and then prove a recurrence relation for $W_n$, which would allow them to find a differential equation~\cite {HassaniEtAl16,Koutschan13a,Koutschan13} and possibly a hypergeometric expression for the Green's function~\cite{Ray14,Guttmann10}.

\subsection {Return Probabilities}

While a random walker on a lattice in one and two dimensions will almost surely return to the origin at some point during its walk, in higher dimensions, this only occurs with some non-trivial probability, which can be expressed in terms of the lattice Green's function:
\begin {equation}
	P = 1 + \frac 1 {\tau \Re {G(\epsilon_\mathrm {min})}} = 1 - \frac 1 {\tau \int_{\epsilon_\mathrm {min}}^{\epsilon_\mathrm {max}} \frac {\rho(\epsilon)} {\epsilon + \tau} \dif \epsilon}. \label {ReturnProbability}
\end {equation}
For example, Watson found the return probability on the three-dimensional face-centered cubic lattice $A_3$ to be~\cite{Watson39}
\begin {equation}
  P_{A_3} = 1 - \frac {16 \sqrt [3] 4 \pi^4} {9 \Gamma(\frac 13)^6}.
\end {equation}
We numerically estimate the return probability of the exceptional lattices to at least three significant figures.

\section{Methods and Results}

We numerically calculate the densities of states using Markov chain Monte Carlo sampling of $\epsilon(\kk)$ over a periodic unit cell.  This entails stochastically jumping between nearby momenta with a probabilistic jump rejection likelihood given in terms of the energies associated with the two momenta by the dispersion relation.  In particular, Metropolis--Hastings weighting~\cite{Hastings70} with asymptotically accurate tails and empirically estimated bulk, combined with proposal jump lengths spanning many orders of magnitude, allows us to obtain results with small relative errors throughout the entire spectrum.  Samples were generated and binned such that the plotted curves are free of relative numerical noise across the entire support.
\begin {figure} [tbh]
	\begin {center}
    	\includegraphics{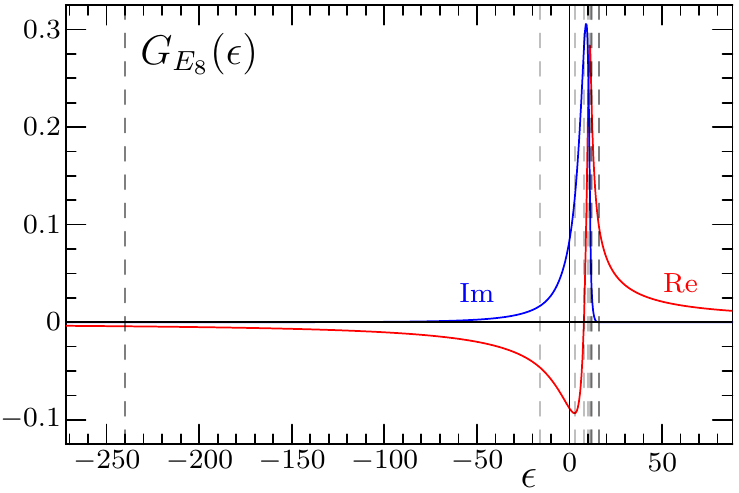}
    \end {center}
        
    \caption {\textbf{$\bm {E_8}$ Lattice Green's Function:} The real and imaginary parts of the $E_8$ lattice Green's function were evaluated using Metropolis--Hasting sampling~\cite {Hastings70}, as detailed in the text, and normalized as in~\cite {Horiguchi72}.  The Van Hove singularities are marked by the vertical dashed lines.  Note that the imaginary part is compactly supported between $\epsilon_\mathrm {min} = -240$ and $\epsilon_\mathrm {max} = 16$.  This imaginary part is plotted as the density of states on a logarithmic scale in figure~\ref {E8DOS}.  The real part can be derived from the imaginary part according to the Kramers--Kronig relation~\eqref {KramersKronig}.} \label {E8LGF}
\end {figure}
The $E_8$ lattice Green's function is shown in figure~\ref {E8LGF}.
\begin {figure} [tbh]
	\begin {center}
    	\includegraphics{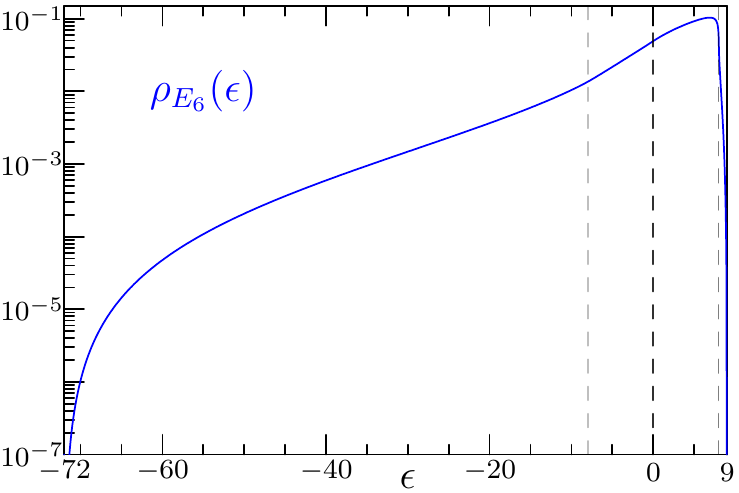}
    \end {center}
     
    \caption {\textbf{$\bm {E_6}$ Density of States:} The Van Hove singularities, enumerated in table~\ref {VanHoves}, are marked by dashed vertical lines, including the bounds of the support at $\epsilon_\mathrm {min} = -72$ and $\epsilon_\mathrm {max} = 9$.  The asymptotic behavior of the tails and the first eight moments of the distribution are given in tables~\ref {Tails} and~\ref {Walks}, respectively.  The real and imaginary parts of the lattice Green's function can be easily calculated from the density of states according to $\Im {G(\epsilon} = \pi \rho(\epsilon)$ and the Kramers--Kronig relation~\eqref {KramersKronig}.} \label {E6DOS}
\end {figure}

\begin {figure} [tbh]
	\begin {center}
    	\includegraphics{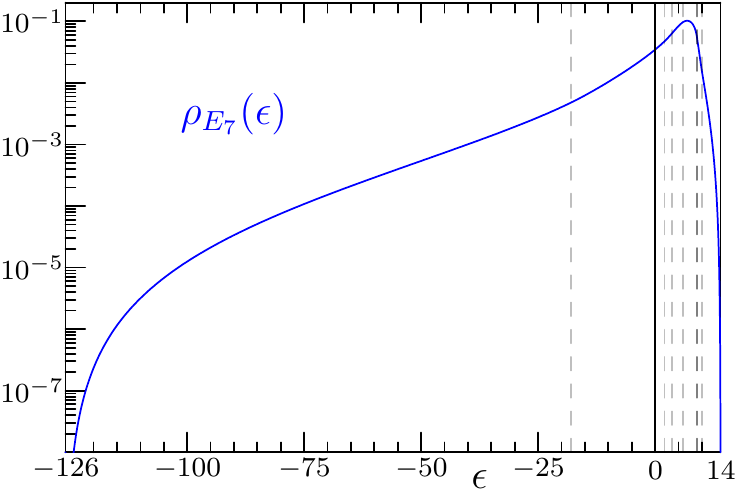}
    \end {center}
        
    \caption {\textbf{$\bm {E_7}$ Density of States:} Equivalent to figure~\ref {E6DOS}, but for $E_7$.  Now $\epsilon_\mathrm {min} = -126$, and $\epsilon_\mathrm {max} = 14$.} \label {E7DOS}
\end {figure}

\begin {figure} [tbh]
	\begin {center}
    	\includegraphics{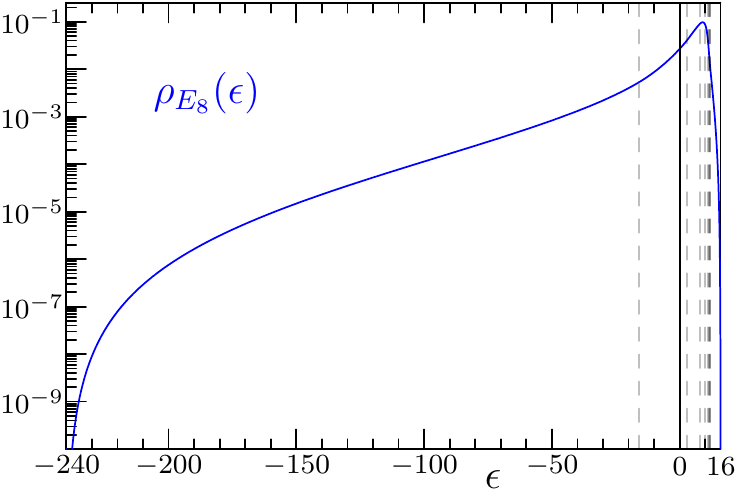}
    \end {center}
        
    \caption {\textbf{$\bm {E_8}$ Density of States:} Equivalent to figures~\ref {E6DOS} and~\ref {E7DOS}, but for $E_8$.  Now $\epsilon_\mathrm {min} = -240$, and $\epsilon_\mathrm {max} = 16$.  The same data, plus a factor of $\pi$, is shown on a linear scale as the imaginary part of figure~\ref {E8LGF}.} \label {E8DOS}
\end {figure}
Because the densities of states contain equivalent information, have compact support, and are nonnegative, we omit the $E_6$ and $E_7$ lattice Green's functions, and instead show only the $E_6$, $E_7$, and $E_8$ densities of states on a logarithmic scale in figures~\ref {E6DOS}--\ref{E8DOS}.

\begin {table} [tbh]
\begin {center}
\begin {tabular} {c|rrrl}
	& \multicolumn{1}{c}{$\epsilon$} & \multicolumn{1}{c}{$n_\downarrow$} & \multicolumn{1}{c}{$n_\uparrow$} \\ \hline
    \multirow{5}{*}{$E_6$} & $-72$ & 0 & 6 & ($\epsilon_\mathrm {min}$) \\
    & $-8$ & 1 & 5 \\
    & 0 & 2 & 4 \\
    & 8 & 1 & 0 & (Degenerate) \\
    & 9 & 6 & 0 & ($\epsilon_\mathrm {max}$) \\ \hline
    \multirow{8}{*}{$E_7$} & $-126$ & 0 & 7 & ($\epsilon_\mathrm {min}$) \\
    & $-18$ & 1 & 6 \\
    & 2 & 1 & 6 \\
    & $18/5 = 3.6$ & 2 & 5 \\
    & 6 & 3 & 4 \\
    & 9 & 2 & 0 & (Degenerate) \\
    & 10 & 6 & 1 \\
    & 14 & 7 & 0 & ($\epsilon_\mathrm {max}$) \\ \hline
    \multirow{11}{*}{$E_8$} & $-240$ & 0 & 8  & ($\epsilon_\mathrm {min}$) \\
    & $-16$ & 1 & 7 \\
    & 3 & 2 & 6 \\
    & 8 & 3 & 5 \\
    & 10 & 4 & 4 \\
    & 11 & 5 & 3 \\
    & $185/16 = 11.5625$ & 6 & 2 \\
    & $320/27 = 11.\overline {851}$ & 7 & 1 \\
    & 12 & 7 & 1 \\
    & 12 & 8 & 0 & (Distinct) \\
    & 16 & 8 & 0 & ($\epsilon_\mathrm {max}$) \\
\end {tabular}
\caption {\textbf {Van Hove Singularities:} The existence and multiplicity of singularities of many indices $n_\downarrow$ is guaranteed by Morse theory~\cite {VanHove}.  They seem to all occur at rational energies.  Perhaps this can be attributed to the symmetry of the lattice?  Note the degenerate critical points at $\epsilon = 8$ and $9$ in $E_6$ and $E_7$ and the existence of two distinct critical points contributing to the singularity at $\epsilon = 12$ in $E_8$.} \label {VanHoves}
\end {center}
\end {table}
The Van Hove singularities, some of which occur at lattice holes, can be determined by the iterated numerical minimization of $\left( \grad_\kk \epsilon(\kk) \right) ^2$, and are enumerated in table~\ref {VanHoves}.  The Van Hove singularities are marked by dashed vertical lines in the figures.  While the usual quadratic critical points cause discontinuities in the $d/2 - 1$\textsuperscript{st} derivative, note that $E_6$ and $E_7$ also exhibit degenerate critical points which lead to stronger singularities.  These are caused by the intersection of isoenergetic planes and lines, respectively, and warrant further investigation.  Curiously, $E_8$ has two distinct Van Hove singularities with an energy of 12, one of which is a local maximum.

Because the global extrema are quadratic, we can evaluate the asymptotic behavior of the density of the state near the tails to be
\begin {equation}
  \rho(\epsilon) \to \frac N V \frac {(2 \pi)^{\frac d 2} |\epsilon - \epsilon_\mathrm {ext.}|^{\frac d 2 - 1}} {\Gamma {\left( \frac d 2 \right) } \sqrt {|H|}}, \label {AsymptoticTails}
\end {equation}
where $N/V$ is the density of the extrema, and $H$ is their Hessian~\eqref {Hessian}.
\begin {table} [tbh]
\begin {center}
\setlength{\tabcolsep}{3pt}
\begin {tabular} {c|r|ccr|lcl}
	& \multicolumn{1}{c|}{$\gamma$} & $\displaystyle \lim_{\epsilon \to \epsilon_\mathrm {min}^+} \frac {\rho_{E_d}(\epsilon)} {(\epsilon - \epsilon_\mathrm {min})^{\frac d 2 - 1}} $ & $\displaystyle \lim_{\epsilon \to \epsilon_\mathrm {max}^-} \frac {\rho_{E_d}(\epsilon)} {(\epsilon_\mathrm {max} - \epsilon)^{\frac d 2 - 1}} $ & $N_\mathrm {max}$ & \multicolumn{3} {c} {$P_{E_d}$ (95\% CI)} \\ \hline \hline
    $E_6$ & 8 & $ \displaystyle \frac {\vphantom{\frac00} 1} {\vphantom{\frac00}2^{13} \cdot 9\sqrt{3} \pi^3} $ & $ \displaystyle \frac 5 {9 \sqrt{3} \pi^3} $ & 80 & 0.022901&--&0.022916 \\ \hline
    $E_7$ & 9 & $ \displaystyle \frac {\vphantom{\frac00} 1} {\vphantom{\frac00}2^7 \cdot 3^8 \cdot 5 \pi^4} $ & $ \displaystyle \frac 3 {160 \pi^4} $ & 36 & 0.011973&--&0.011982 \\ \hline
    $E_8$ & 15 & $ \displaystyle \frac {\vphantom{\frac00} 1} {\vphantom{\frac00}2^{13} \cdot 3^5 \cdot 5^4 \pi^4} $ & $ \displaystyle \frac {45} {2^{13} \pi^4} $ & 135 & 0.0059014&--&0.0059064
\end {tabular}
\caption {\textbf {Asymptotic Tails and Return Probabilities:} First, the $\gamma$ skewness/extrema imbalance, $-\epsilon_\mathrm {min}/\epsilon_\mathrm {max}$, which is important for the stability of the corresponding crystalline spatial structures~\cite {Savitz17}, is given.  The asymptotic behavior of the tails of the densities of states follows, evaluated from the multiplicity and Hessian determinant of the extrema according to equation~\eqref {AsymptoticTails}.  $N_\mathrm {min}$ is always unity, corresponding to the lattice points.  The 95\% confidence intervals for the return probability in the final column were calculated by evaluating $\overline \epsilon/\epsilon_\mathrm {min}$ with the relative Metropolis--Hasting weighting $1/ \left( \epsilon - \epsilon_\mathrm {min} \right) $ over $10^{11}$ momenta.} \label {Tails}
\end {center}
\end {table}
The results of this calculation are given in table~\ref {Tails}.  95\% confidence intervals for the return probabilities are also included, calculated according to $P = \overline \epsilon/\epsilon_\mathrm {min}$, where $\overline \epsilon$ is the average value of $\epsilon$ with the relative Metropolis--Hasting weighting $1/ \left( \epsilon - \epsilon_\mathrm {min} \right) $, which is equivalent to equation~\eqref {ReturnProbability}, and sampled over $10^{11}$ momenta.

\begin {table} [tbh]
\begin {center}
\setlength{\tabcolsep}{2pt}
\begin {tabular} {c|r!{\color{gray}\vrule}r!{\color{gray}\vrule}r!{\color{gray}\vrule}r!{\color{gray}\vrule}r!{\color{gray}\vrule}r|l}
	& \multicolumn{1}{c}{$W_3$} & \multicolumn{1}{c}{$W_4$} & \multicolumn{1}{c}{$W_5$} & \multicolumn{1}{c}{$W_6$} & \multicolumn{1}{c}{$W_7$} & \multicolumn{1}{c|}{$W_8$} & \multicolumn{1}{c}{OEIS \#} \\ \hline
    $E_6$ & 1\,440 & 54\,216 & 2\,134\,080 & 93\,993\,120 & 4\,423\,628\,160 & 219\,463\,602\,120 & \href {http://oeis.org/A292881} {A292881} \\
    $E_7$ & 4\,032 & 228\,690 & 14\,394\,240 & 1\,020\,623\,940 & 78\,353\,170\,560 & 6\,393\,827\,197\,170 & \href {http://oeis.org/A292882} {A292882} \\
    $E_8$ & 13\,440 & 1\,260\,720 & 137\,813\,760 & 17\,141\,798\,400 & 2\,336\,327\,078\,400 & 341\,350\,907\,713\,200 & \href {http://oeis.org/A292883} {A292883}
\end {tabular}
\caption {\textbf {Closed Walk Counts:} In each case, $W_0 = 1$, $W_1 = 0$, and $W_2 = \tau$, the kissing number of the lattice given in section~\ref{ADE}.  These subsequent closed walk counts were evaluated by brute force enumeration of the walks, using the symmetry of the lattices to fix the first and last steps.  These sequences were uploaded to the On-Line Encyclopedia of Integer Sequences as \href {http://oeis.org/A292881} {A292881}--\href {http://oeis.org/A292883} {A292883}~\cite {OEIS}.  They give the moments of the densities of states distributions, as detailed in equation~\eqref {GeneratingFunction}.  More powerful techniques should be devised to evaluate many further terms, which would allow an ``experimental'' approach like that in~\cite{Koutschan13,Zenine15,HassaniEtAl16} to find analytic results for the lattice Green's functions.} \label {Walks}
\end {center}
\end {table}
The number of closed walks up to eight steps in length were computed by brute force, using symmetry to fix the first and last steps, and are given in table~\ref{Walks}.  These sequences have been listed on the On-Line Encyclopedia of Integer Sequences as \href {http://oeis.org/A292881} {A292881}, \href {http://oeis.org/A292882} {A292882}, and \href {http://oeis.org/A292883} {A292883}~\cite {OEIS}.

\section{Discussion}

The numerous and irregular Van Hove singularities are indicative of the rich band structure exhibited by these lattices.  Varying the Fermi energy through one of these singularities leads to a topology change of the Fermi sea in momentum space.   It is curious that the Van Hove singularities all occur at rational energies.  Why is this?  Their existence and multiplicity are ensured by Morse theory~\cite {VanHove} and the symmetries of the lattice.  Perhaps similar arguments can be found to explain the energies' rationality.  The degenerate critical points at the intersection of isoenergetic momentum planes and lines in the band structure $\epsilon(\kk)$ of $E_6$ and $E_7$ at $\epsilon = 8$ and $9$ respectively are interesting and deserve further investigation.  Oddly, the singularity at $\epsilon = 12$ in $E_8$ consists of critical points of two distinct signatures, including local maxima, which are somewhat analogous to maxons~\cite{BennemannKetterson78}.  Given the high expected energetic stability of the exceptional lattices~\cite {Savitz17}, one must wonder whether \mbox {six-,} \mbox{seven-,} and eight-spatial dimensional physicists with metals analogous to our own would be intimately familiar with the complexities of these band structures.

The ratios of the ground and highest excited states' energies, a metric which is important for assessing the stability of the corresponding crystalline spatial structure in the Barkan--Diamant--Lifshitz model~\cite{BDL,Savitz17}, are $\gamma_{E_6} = -\epsilon_\mathrm {min}/\epsilon_\mathrm {max} = 8$, $\gamma_{E_7} = 9$, and $\gamma_{E_8} = 15$.  For the non-exceptional infinite families $A_d$ and $D_d$, it appears that $\gamma_{A_d} = d$, $\gamma_{D_{2d}} = 2d - 1$, and $\gamma_{D_{2d + 1}} = 2d + 1$.

We encourage the reader to find more powerful approaches for the enumeration of the closed walks given in table~\ref {Walks}, such as the ``creative telescoping'' Koutschan \emph {et al.}~has applied to the face-centered cubic lattices, as sufficient data and/or a recurrence relation could lead to differential equations for the Green's functions~\cite{Koutschan13,Zenine15,HassaniEtAl16} or even closed-form hypergeometric expressions~\cite{Ray14,Guttmann10}.

It would be interesting to extend this analysis to the infinite $A_d$ and $D_d$ lattice families, non-root lattices, and further quasilattices~\cite {Savitz17}. Another enticing target is the twenty-four dimensional Leech lattice~\cite {Leech64,Leech67,Conway69}, which has been proven to be the optimal sphere packing in twenty-four dimensions~\cite{CohnEtAl17} and plays a key role in monstrous moonshine~\cite{FrenkelLepowskyMeurman}. We hope to apply these results to the evaluation of certain arbitrarily high-dimensional crystal energies in the Barkan--Diamant--Lifshitz~\cite {BDL} model in future work.

\begin{acknowledgement}
  Thanks to Gil Refael, Yen Lee Loh, Christoph Koutschan, Evert van Nieuwenburg, Matthew Heydeman, and Petr Kravchuk for useful discussions and encouragement.  This work was supported by the Institute of Quantum Information and Matter, a National Science Foundation frontier center partially funded by the Gordon and Betty Moore Foundation, the Packard Foundation and the National Science Foundation through award DMR-1410435.  M.B.~thanks the Caltech Student--Faculty Programs office and Marcella Bonsall for their support.
\end{acknowledgement}

\end{document}